\documentclass[letterpaper,twocolumn]{jpsj3}
\usepackage{graphicx}
\usepackage{color}

\begin{document}

\title{Quantum Singular Value Decomposition of Spin Correlation Matrix in One-Dimensional Heisenberg Model}
\author{Kohei Ohgane, Tatsuya Kumamoto, and Hiroaki Matsueda\thanks{hiroaki.matsueda.c8@tohoku.ac.jp}}
\inst{
Department of Applied Physics, Graduate School of Engineering, Tohoku University, Sendai 980-8579, Japan
}
\recdate{\today}

\abst{
We present singular value decomposition of spin correlation matrix defined from the ground state of one-dimensional antiferromagnetic quantum Heisenberg model. The decomposition creates a data set that coinsides with various domain excitations from classical antiferromagnetic state. We determine the scaling relation for the singular values as a function of the domain size. The singular values are closely related to the square of weights of the bases in the ground-state wavefunction. The nature of the singular value decomposition is to precisely estimate the appropriate bases and corresponding weights of the ground-state wavefunction.
}

\maketitle

\section{Introduction}

Singular value decomposition (SVD) is a powerful tool for principle component analysis (PCA). Historically, SVD has been applied to wide area beyond simple data analysis. For example, in condensed matter physics, SVD is a core algorithm of density matrix renormalization group (DMRG)~\cite{White1,White2}, which provides us with a systematic way of doing quantum simulation on interacting one-dimensional (1D) lattice models. In DMRG calculation, automatic truncation of unimportant states makes it possible to precisely treat hundreds of lattice sites, and it has been shown that the truncation is closely related to how precise we treat quantum entanglement inherent in our target quantum models~\cite{Cirac}.

Recently, a quantum version of SVD attracts much attention because of development of quantum computer technology~\cite{Lloyd1,Mosetti,Lloyd2,Tang}. In data science community, an important request is that people can carry out standard (classical) SVD for massive data efficiently by high performance quantum computer. The algorithm to realize it is sometimes called quantum SVD. Similarly, quantum speed up of neural network and machine learning is called quantum neural network and quantum machine learning. Even in classical problems, there are many NP hard problems, and thus this strategy arises from current data science and social application quite naturally.

In contrast, our strategy in condensed matter physics side is to apply PCA to quantum data directly in order to extract essential information of our target quantum many-body systems. This is because the Hilbert spaces of our target models are exponentially large and only limited simulations are possible even if we use super-parallel computers. In that sense the real efficient quantum PCA is necessary in condensed matter physics side. Many trials based on tensor-network variational optimization for the ground state are milestones toward the ultimate success of those simulations~\cite{Ostlund1,Ostlund2,Verstraete1,Verstraete2,Vidal,Evenbly1}. 

We simply imagine that the first component of the best PCA for the ground state represents a macroscopic state and in special cases the state would be an ordered state in the classical limit. Then, the higher-order components of PCA systemetically include quantum fluctuation with various length scales. Even if the strong quantum fluctuation breaks this picture, it is still meaningful to consider how the weights of the higher-order components dominate. This imagination might be too naive, since there are some cases, like topological order, that have no classical correspondence. However, we cannot throw the authentic ideas away. It is because the viewpoint of order and various-scale fluctuations are quite natural for human recognition, and this natural sense is a core of PCA. Unfortunately, the algorithms of DMRG and tensor-network methods do not follow this consideration. In these methods, it is very hard to treat entanglement throughout the superblock (whole system) directly, when we propose realistic numerical algorithms. What we can do everytime is sequential optimization of local parts of the whole wavefunction. For the realization of desired quantum PCA, it is worth mentioning to find a quantity that reflects nonlocal entanglement throughout the system. Since the entanglement entropy is roughly given by the logarithm of a particular kind of two-point correlator in terms of conformal field theory, the quantity we need should be associated with the correlation function. For quantum many-body systems, the direct connection between entanglement and correlation function is still unclear, even though the study of entanglement is a recent great trend in many branches of physics. Thus, our examination contributes to the development of entanglement approach to condensed matter physics.

A SVD analysis of classical spin models has been performed previously by one of the present authors (HM)~\cite{Matsueda1,Matsueda2,Okunishi1,Matsueda3,Ozaki1,Ozaki2}. In this analysis, a spin snapshot created by Monte Carlo simulation was regarded as a matrix. In particular, the snapshot at the critical point was decomposed into a set of patterns with different cluster sizes. Due to the nature of definition of SVD, the SVD spectrum for the snapshot data represents H$\ddot{\rm o}$lder conjugate of two-point spin correlator. Then, we can pick up a critical exponent from one snapshot near the critical point, and need not to treat whole information of partition function. Therefore, the length-scale decomposition mechanism of SVD seems to fit with the present purpose. Unfortunately, in quantum spin models, the superposition of states makes it useless to introduce a definite spin pattern. If we overcome this difficulty, we can extend this unique method to various quantum systems.

Motivated by quantum SVD and the previous works, we focus on the quantum correlation matrix of the 1D Heisenberg model, and analyze it by SVD. We will find that the SVD of the correlation matrix can precisely estimate information of the appropriate bases and the corresponding weights of the ground-state wavefunction. The decomposition naturally creates a hierarchy of data set that coinsides with various domain excitations from the classical antiferromagnetic spin state. For these excitations as appropriate bases of the ground-state wavefunction, the corresponding weights are determined from the SVD spectrum of the spin correlation matrix. Therefore, instead of considering local entanglement, we can define PCA that starts with the macroscopic order and systematically introduces correction. It is thus necessary to introduce the SVD of the correlation matrix for desired quantum PCA.

The outline of this paper is as follows. In the next section we define our model, physical quantity we are going to focus on, and SVD method. In the third section, we perform detailed analysis based on exact diagonalization form small clusters and numerical optimization based on matrix product state (MPS) for larger systems. In Sec.4, we discuss implications of the present results in terms of circulant matrix approach. In the final section, we summarize our work, and comment on future perspectives.

\section{Model and Method}

Let us start with the antiferromagnetic Heisenberg Hamiltonian in spatially one dimension:
\begin{eqnarray}
H=J\sum_{i=1}^{N}\vec{S}_{i}\cdot\vec{S}_{i+1}.
\end{eqnarray}
Here $\vec{S}$ is quantum spin operator, $\vec{S}=\frac{1}{2}\vec{\sigma}$ with Pauli matrices $\vec{\sigma}$, $N$ is the number of lattice sites (we assume that $N$ is a power of $2$), and we assume the periodic boundary condition, $\vec{S}_{N+1}=\vec{S}_{1}$. The ground state of this Hamiltonian is represented as $\left|\psi\right>$.

In this paper, we calculate a set of all possible two-point spin correlators at zero temperature
\begin{eqnarray}
S_{ij}=\left<\psi\right|S_{i}^{z}S_{j}^{z}\left|\psi\right>=\frac{1}{3}\left<\psi\right|\vec{S}_{i}\cdot\vec{S}_{j}\left|\psi\right>,
\end{eqnarray}
and construct the following correlation matrix
\begin{eqnarray}
S=\left(\begin{array}{cccc}S_{11}&S_{12}&\cdots&S_{1N}\\S_{21}&S_{22}&\cdots&S_{2N}\\ \vdots&\vdots&&\vdots\\S_{N1}&S_{N2}&\cdots&S_{NN}\end{array}\right).
\end{eqnarray}
We apply SVD to this matrix. Unfortunately, the evaluation of all entries is not easy in general. However, once we obtain those values, the SVD is very easily done even for large $N$ cases. In this paper, we focus on the analysis of functionality of the SVD decomposition of $S$. In the previous SVD work for classical models~\cite{Matsueda1,Matsueda2,Okunishi1,Matsueda3,Ozaki1,Ozaki2}, we considered a particular spin pattern generated by Monte Carlo method, and regarded it as a matrix. In the quantum case, however, it is impossible to define a particular spin pattern due to quantum superposition. Instead of using snapshots, we introduce the abovementioned correlation matrix to treat all possible quantum correlation.

Note that this type of correlator matrix was previously used for evaluation of entanglement Hamiltonian for free fermions~\cite{Henley1,Henley2}. Furthermore, the correlation matrix is also considered for definition of space-time metric in the exact holographic mapping~\cite{XiaoLiang,ChingHua}. The holography is a complementary concept to the entanglement. Thus, the correlation matrix is a key quantity to consider entanglement and holography. In these previous works, SVD was not examined yet, even though SVD is a key method to extract essential information of our target system. Therefore, we would like to clarify the functionality of SVD to quantum correlation matrix.

The SVD of the correlation matrix is defined by
\begin{eqnarray}
S_{ij}=\sum_{n=1}^{N}U_{in}\sqrt{\lambda_{n}}V_{jn},
\end{eqnarray}
where $\sqrt{\lambda_{n}}$ are singular values and $U_{in}$ and $V_{jn}$ are column unitary matrices. Because of the symmetry $S_{ij}=S_{ji}$, we find $U_{in}=V_{jn}$. These conditions show that SVD is equivalent to the matrix diagonalization in the present case and $\sqrt{\lambda_{n}}$ is the eigenvalue of the matrix $S$. For this reason, we obtain the following identity
\begin{eqnarray}
\sum_{n=1}^{N}\sqrt{\lambda_{n}}=tr S=tr\Lambda=\frac{N}{4}, \label{N4}
\end{eqnarray}
where $\Lambda$ is the diagonal matrix with $\Lambda_{nn}=\sqrt{\lambda_{n}}$.

Our correlation matrix $S$ is a circulant matrix that is a special version of the Toeplitz matrix. This matrix has beautiful properties associated with periodicity of the model. Let us represent the correlation matrix as
\begin{eqnarray}
S=\left(\begin{array}{ccccc}s_{0}&s_{N-1}&\cdots&s_{2}&s_{1}\\s_{1}&s_{0}&s_{N-1}&&s_{2}\\ \vdots&s_{1}&s_{0}&\ddots&\vdots\\s_{N-2}&&\ddots&\ddots&s_{N-1}\\s_{N-1}&s_{N-2}&\cdots&s_{1}&s_{0}\end{array}\right),
\end{eqnarray}
where $s_{0}=1/4$ for arbitrary $N$ value. Due to the periodic boundary condition, we have an additional condition
\begin{eqnarray}
s_{j}=s_{N-j}, \label{sscondition}
\end{eqnarray}
for $j=1,2,...,N/2-1$. This condition is also related to the Hermitian condition of $S$. The eigenvectors of $S$ (it is not necessary to assume Eq.~(\ref{sscondition})) are given by
\begin{eqnarray}
\vec{X}_{k}=\frac{1}{\sqrt{N}}\left(\begin{array}{c}1\\ \omega_{k}\\ \omega_{k}^{2}\\ \vdots \\ \omega_{k}^{N-1} \end{array}\right), \label{Xvector}
\end{eqnarray}
where $\omega_{k}=\left(e^{2\pi i/N}\right)^{k}$ for $k=1,2,...,N$, $\omega_{1}$ is the $N$-th root of $1$, and the eigenvectors are columns of the discrete Fourier transformation matrix. The corresponding eigenvalues are given by
\begin{eqnarray}
\chi_{k}=s_{0}+s_{N-1}\omega_{k}+s_{N-2}\omega_{k}^{2}+\cdots +s_{1}\omega_{k}^{N-1}.
\end{eqnarray}
A striking fact is that all the information of the correlation functions are contained in the eigenvalues. Therefore, the SVD spectrum, not the basis, is a key quantity for quantum PCA. Furthermore, all the eigenvalues are real due to the presence of Eq.~(\ref{sscondition}). We find the following compact form
\begin{eqnarray}
\chi_{k}=s_{0}+s_{N/2}(-1)^{k}+2\sum_{m=1}^{N/2-1}s_{m}c_{mk},
\end{eqnarray}
where
\begin{eqnarray}
c_{mk}=\cos\left(\frac{2\pi mk}{N}\right). \label{cmk}
\end{eqnarray}
We have $\sum_{k=1}^{N}\chi_{k}=N/4$ from Eq.~(\ref{N4}), and this is equal to $\sum_{m}\sum_{k}s_{m}c_{mk}=0$. Note that the label $k$ does not indicate descending order of $\chi_{k}$. We need to change the order for the analysis of the singular value distribution. According to the nature of $s_{m}$ in the antiferromagnetic system, we find the following descending order as well as characteristic degenerate structure
\begin{eqnarray}
\chi_{N/2}&>&\chi_{N/2-1}=\chi_{N/2+1}>\chi_{N/2-2}=\chi_{N/2+2} \nonumber \\
&>&\cdots>\chi_{1}=\chi_{N-1}>\chi_{N}.
\end{eqnarray}
This is a very important relation throughout this paper. We will find $\chi_{N}=0$, and it leads to $\sum_{m=0}^{N-1}s_{m}=0$. The relationship with the singular values is given by
\begin{eqnarray}
\sqrt{\lambda_{1}}=\chi_{N/2} \; , \; \sqrt{\lambda_{2m}}=\sqrt{\lambda_{2m+1}}=\chi_{N/2\pm m},
\end{eqnarray}
for $m=2,4,6,...$. The relationship between the indices of $\lambda_{n}$ and $\chi_{k}$ is thus given by
\begin{eqnarray}
n=N-2k, \label{nktrans}
\end{eqnarray}
for $k=1,2,...,N/2-1$. In the following, we first try to examine a small system exactly, and then perform numerical simulation based on MPS for larger systems. On the basis of the Fourier transformation as well as the singular value spectrum as a key quantity for PCA, we will examine how to extract wavefunction information from the correlation data.

\section{Result}

\subsection{Preliminary: Exact analysis for 4-site ring}

Let us first consider the 4-site case in which we can obtain the exact eigenstates. The ground state for $S_{tot}^{z}=0$ is a resonant state of spin singlet pair:
\begin{eqnarray}
\left|\psi\right>&=&\frac{1}{\sqrt{12}}\left(-\left|\uparrow\uparrow\downarrow\downarrow\right>+2\left|\uparrow\downarrow\uparrow\downarrow\right>-\left|\uparrow\downarrow\downarrow\uparrow\right> \right. \nonumber \\
&&\left.-\left|\downarrow\uparrow\uparrow\downarrow\right>+2\left|\downarrow\uparrow\downarrow\uparrow\right>-\left|\downarrow\downarrow\uparrow\uparrow\right>\right) \\
&=& \frac{1}{\sqrt{12}}\left|\uparrow\downarrow-\downarrow\uparrow\right>_{12}\otimes\left|\uparrow\downarrow-\downarrow\uparrow\right>_{34} \nonumber \\
&&+\frac{1}{\sqrt{12}}\left|\uparrow\downarrow-\downarrow\uparrow\right>_{41}\otimes\left|\uparrow\downarrow-\downarrow\uparrow\right>_{23}. \label{RVB}
\end{eqnarray}

Before going into evaluation of quantum correlation matrix, we summarize the entanglement properties of this state, and point out some disadvantage of using DMRG and multiscale entanglement renormalization ansatz (MERA)~\cite{Vidal}. To define entanglement, we need to divide whole system into two (subsystem and environment), and various patterns of spatial division exist. We call site 1 (2) as $A$ ($B$). The partial density matrix for subsystem $A\otimes B$ is defined by
\begin{eqnarray}
\rho_{A\otimes B}&=&tr_{\overline{A\otimes B}}\left|\psi\right>\left<\psi\right| \nonumber \\
&=&\frac{1}{12}\left|\uparrow\uparrow\right>\left<\uparrow\uparrow\right|+\frac{1}{12}\left|\downarrow\downarrow\right>\left<\downarrow\downarrow\right| \nonumber \\
&& +\frac{5}{12}\left|\uparrow\downarrow\right>\left<\uparrow\downarrow\right|+\frac{5}{12}\left|\downarrow\uparrow\right>\left<\downarrow\uparrow\right| \nonumber \\
&& -\frac{1}{3}\left|\uparrow\downarrow\right>\left<\downarrow\uparrow\right|-\frac{1}{3}\left|\downarrow\uparrow\right>\left<\uparrow\downarrow\right|,
\end{eqnarray}
and the entanglement entropy is evaluated as
\begin{eqnarray}
S_{A\otimes B}=-tr_{A\otimes B}\left(\rho_{A\otimes B}\log\rho_{A\otimes B}\right)=2\log 2 -\frac{1}{2}\log 3.
\end{eqnarray}
Similarly, we calculate the partial density matrix for single site $A$
\begin{eqnarray}
\rho_{A}=tr_{\overline{A}}\left|\psi\right>\left<\psi\right|=\frac{1}{2}\left|\uparrow\right>\left<\uparrow\right|+\frac{1}{2}\left|\downarrow\right>\left<\downarrow\right|
\end{eqnarray}
and the entanglement entropy is evaluated as
\begin{eqnarray}
S_{A}=-tr_{A}\left(\rho_{A}\log\rho_{A}\right)=\log 2.
\end{eqnarray}
We also obtain
\begin{eqnarray}
S_{B}=-tr_{B}\left(\rho_{B}\log\rho_{B}\right)=\log 2,
\end{eqnarray}
where $\rho_{B}=tr_{\overline{B}}\left|\psi\right>\left<\psi\right|$. The mutual information is also given by
\begin{eqnarray}
I_{M}=S_{A}+S_{B}-S_{A\otimes B}=\frac{1}{2}\log 3.
\end{eqnarray}
These entropy values characterize the formation of fluctuation of singlet pairs represented in Eq.~(\ref{RVB}). In particular, the leading term of the magnitude of $S_{A\otimes B}$, $2\log 2$, represents two singlets (the second term in Eq.~(\ref{RVB})) which are cut by partial truncation of environmental degrees of freedom, $\overline{A\otimes B}$. The mutual information represents net quantum correlation between $A$ and $B$ embedded into whole $4$ site ring. These entropies seem to be very nice parameters to estimate the entanglement structure of the model. On the other hand, when we look at the eigenstates of partial density matrix $\rho_{A\otimes B}$, the situation changes. We easily find that the triplet eigenstates for the eigenvalue $1/12$ are $\left|\uparrow\uparrow\right>$, $\left|\downarrow\downarrow\right>$, and $\left|\uparrow\downarrow+\downarrow\uparrow\right>/\sqrt{2}$, and the singlet eigenstate for the eigenvalue $9/12$ is $\left|\uparrow\downarrow-\downarrow\uparrow\right>/\sqrt{2}$. The leading component is singlet, but this singlet is located inside of partial system. The nonlocal entanglement across the boundary must be represented by complex combination of triplet states. They are very asymmetric. Thus, the two terms in Eq.~(\ref{RVB}) are not equally treated by a simple SVD for the wavefunction approximation. In DMRG calculation, this asymmetry is a reason for worse numerical convergence in periodic boundary condition. In the MERA tensor network, this asymmetric nature is somehow relaxed by the introduction of disentangler tensors. However, this relaxation is imperfect. For the ground state $\left|\psi\right>=\sum_{s_{1},...,s_{4}}\psi^{s_{1}...s_{4}}\left|s_{1}...s_{4}\right>$, the MERA corresponds decomposition of the coefficient $\psi^{s_{1}...s_{4}}$ by a set of functional tensors:
\begin{eqnarray}
\psi^{s_{1}...s_{4}}=\sum_{a,b}\sum_{\alpha,\beta,\gamma,\delta}T^{ab}W_{a}^{\delta\alpha}W_{b}^{\beta\gamma}U^{s_{2}s_{3}}_{\alpha\beta}U^{s_{4}s_{1}}_{\gamma\delta}.
\end{eqnarray}
In this representation, the two types of singlet pairs in Eq.~(\ref{RVB}) is still treated asymmetrically. To relax this asymmetry, we need to introduce extra tensor dimension, or combination of different types of networks that are consistent with all terms in Eq.~(\ref{RVB}).

To automatically find correct information of the wavefunction by overcoming the abovementioned weak points, it is efficient for us to introduce SVD of quantum correlation matrix. In the present approach, the asymmetric treatment of singlet pairs does not occur, and the successful quantum PCA is possible as we will see later.

On the basis of the abovementioned facts, we analyze the correlation matrix which can be evaluated as
\begin{eqnarray}
S=\left(\begin{array}{cccc}\frac{1}{4}&-\frac{1}{6}&\frac{1}{12}&-\frac{1}{6}\\ -\frac{1}{6}&\frac{1}{4}&-\frac{1}{6}&\frac{1}{12}\\ \frac{1}{12}&-\frac{1}{6}&\frac{1}{4}&-\frac{1}{6}\\ -\frac{1}{6}&\frac{1}{12}&-\frac{1}{6}&\frac{1}{4}\end{array}\right).
\end{eqnarray}
The matrix $S$ is decomposed into SVD components as
\begin{eqnarray}
S=\sum_{n=1}^{4}S^{(n)},
\end{eqnarray}
and
\begin{eqnarray}
\left(S^{(n)}\right)_{ij}=U_{in}\sqrt{\lambda_{n}}V_{jn},
\end{eqnarray}
where the singular values are given by
\begin{eqnarray}
\sqrt{\lambda_{1}}=\frac{2}{3} \; , \; \sqrt{\lambda_{2}}=\sqrt{\lambda_{3}}=\frac{1}{6} \; , \; \lambda_{4}=0 , \label{lambdan}
\end{eqnarray}
and we confirm $\sqrt{\lambda_{1}}+\sqrt{\lambda_{2}}+\sqrt{\lambda_{3}}=1$. All non-zero components of SVD are 
\begin{eqnarray}
S^{(1)}=\frac{1}{6}\left(\begin{array}{cccc}1&-1&1&-1\\-1&1&-1&1\\1&-1&1&-1\\-1&1&-1&1\end{array}\right),
\end{eqnarray}
\begin{eqnarray}
S^{(2)}=\frac{1}{12}\left(\begin{array}{cccc}1&0&-1&0\\0&0&0&0\\-1&0&1&0\\0&0&0&0\end{array}\right),
\end{eqnarray}
and
\begin{eqnarray}
S^{(3)}=\frac{1}{12}\left(\begin{array}{cccc}0&0&0&0\\0&1&0&-1\\0&0&0&0\\0&-1&0&1\end{array}\right).
\end{eqnarray}
We find that the first principle component, $S^{(1)}$, clearly represents the N\'{e}el order, the classical limit of our ground state. The higher-order components, $S^{(2)}$ and $S^{(3)}$, represent some magnetic excitation from the N\'{e}el order. Note that the second and the third components share the same singular value.

Going back to the circulant matrix approach, we find
\begin{eqnarray}
S^{(1)}=\chi_{2}\vec{X}_{2}\vec{X}_{2}^{\dagger},
\end{eqnarray}
and
\begin{eqnarray}
S^{(2)}+S^{(3)}=\chi_{1}\vec{X}_{1}\vec{X}_{1}^{\dagger}+\chi_{3}\vec{X}_{3}\vec{X}_{3}^{\dagger},
\end{eqnarray}
where $\chi_{2}=\sqrt{\lambda_{1}}=2/3$, $\chi_{1}=\chi_{3}=\sqrt{\lambda_{2}}=\sqrt{\lambda_{3}}=1/6$, and $\chi_{4}=0$. Note that the each doubly-degenerate component in the circulant matrix approach, $\vec{X}_{1}\vec{X}_{1}^{\dagger}$ and $\vec{X}_{3}\vec{X}_{3}^{\dagger}$, is different from $S^{(2)}$ and $S^{(3)}$ and contains complex entries
\begin{eqnarray}
\chi_{1}\vec{X}_{1}\vec{X}_{1}^{\dagger}=\frac{1}{24}\left(\begin{array}{cccc}1&-i&-1&i\\i&1&-i&-1\\-1&i&1&-i\\-i&-1&i&1\end{array}\right) , \\
\chi_{3}\vec{X}_{3}\vec{X}_{3}^{\dagger}=\frac{1}{24}\left(\begin{array}{cccc}1&i&-1&-i\\-i&1&i&-1\\-1&-i&1&i\\i&-1&-i&1\end{array}\right) ,
\end{eqnarray}
and the complex entries vanish completely after summation of degenerate states
\begin{eqnarray}
\chi_{1}\vec{X}_{1}\vec{X}_{1}^{\dagger}+\chi_{3}\vec{X}_{3}\vec{X}_{3}^{\dagger}=\frac{1}{12}\left(\begin{array}{cccc}1&0&-1&0\\0&1&0&-1\\-1&0&1&0\\0&-1&0&1\end{array}\right) .
\end{eqnarray}
To aviod ambiguity of representation between the standard SVD and circulant matrix approaches, we should notice that only the sum of the degenerate states is a physically relevant quantity. 

In terms of Fourier analysis, the principle component, $S^{(1)}$, has periodicity characterized by the wave number $\pi$. The sum of second and third components, $S^{(2)}+S^{(3)}$, can be characterized by the wave number $\pi/2$. These wave numbers are consistent with the cosine factor in Eq.~(\ref{cmk}). We would like to know much clearer physical meaning of this wave number. The wave number is related to the size of the antiferromagnetic domain. To see this feature we redefine the wavefunction as 
\begin{eqnarray}
\left|\psi\right>&=&\left|\psi_{1}\right>+\left|\psi_{2}\right>, \\
\left|\psi_{1}\right>&=&\frac{1}{\sqrt{3}}\left(\left|\uparrow\downarrow\uparrow\downarrow\right>+\left|\downarrow\uparrow\downarrow\uparrow\right>\right), \\
\left|\psi_{2}\right>&=&-\frac{1}{\sqrt{12}}\left(\left|\uparrow\uparrow\downarrow\downarrow\right>+\left|\uparrow\downarrow\downarrow\uparrow\right>+\left|\downarrow\downarrow\uparrow\uparrow\right>+\left|\downarrow\uparrow\uparrow\downarrow\right>\right). \nonumber \\
&&
\end{eqnarray}
The first component $\left|\psi_{1}\right>$ is nothing but the sum of classical antiferromagnetic spin configurations. Due to the periodic boundary condition, the second component $\left|\psi_{2}\right>$ represents a combination of one domain excitation (two domain walls) from the classical antiferromagnetic state. For instance, the state $\left|\uparrow\downarrow\downarrow\uparrow\right>$ has two domains $\uparrow\downarrow$ and $\downarrow\uparrow$ from left to right, and this is realized by one domain flip from the pure antiferromagnetic state. Let us introduce matrices $S_{n}$ with $n=1,2$ whose entries are defined by
\begin{eqnarray}
\left(S_{n}\right)_{ij}=\left<\psi_{n}\right|S_{i}^{z}S_{j}^{z}\left|\psi_{n}\right>.
\end{eqnarray}
Then we find
\begin{eqnarray}
S_{1} &=& S^{(1)}, \\
S_{2} &=& S^{(2)}+S^{(3)}.
\end{eqnarray}
Therefore, the SVD component for the quantum correlation matrix is nothing but the quantity that reflects precise information of the ground-state wavefunction. Furthermore, when we rewrite the ground state as $\left|\psi_{i}\right>=c_{i}\left|\phi_{i}\right>$ with $c_{1}=1/\sqrt{3}$ and $c_{2}=-1/\sqrt{12}$, we find $\sqrt{\lambda_{1}}=2c_{1}^{2}$ and $\sqrt{\lambda_{2}}=2c_{2}^{2}$. The interesting point is that the SVD for the quantum correlation matrix does not introduce the concept of entanglement and more classical view like domain excitations is essential for understanding the wavefunction structure.

\subsection{Numerical results}

To examine the nature of quantum correlation matrix for larger systems, we perform numerical calculation based on MPS optimization of the ground state on lattice with $N=64$ under the periodic boundary condition. We assume the form of MPS as
\begin{eqnarray}
\left|\psi\right>=\sum_{s_{1}=\uparrow,\downarrow}\cdots\sum_{s_{N}=\uparrow,\downarrow}tr\left(A_{1}^{s_{1}}\cdots A_{N}^{s_{N}}\right)\left|s_{1}\cdots s_{N}\right>,
\end{eqnarray}
and optimize it so that the variational energy, $E=\left<\psi\right|H\left|\psi\right>/\left<\psi|\psi\right>$, is minimized. For the minimization, we solved the generalized eigenvalue problem $H_{\rm eff}\left|A\right>=EN_{\rm eff}\left|A\right>$ for the vector $\left|A\right>$ that is defined by one-dimensional arrangement of each matrix. Here, the dimension of each matrix $A_{i}^{s_{i}}$, $\chi$, is taken to be up to $\chi=10$. Starting from random matrices, we repeat optimization of all matrices sequentially by $40$ times. The energy eigenvalue almost converges by much smaller number of iterations (typically $2$ times), but we need many iterations for convergence of the singular value spectrum. We also perform exact diagonalization calculation for $N=12$ in order to confirm reliability of the MPS calculation (not shown here).

\begin{figure}[htbp]
\begin{center}
\includegraphics[width=6cm]{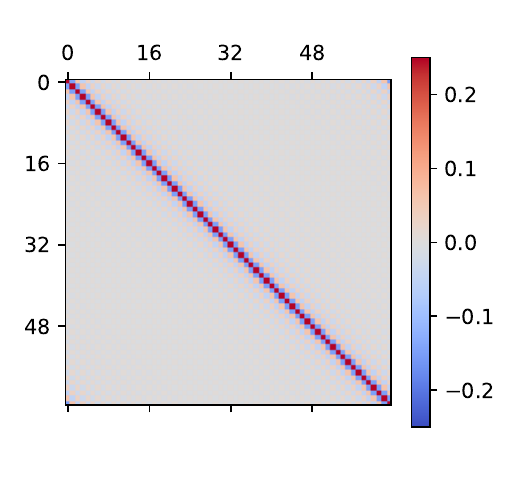}
\end{center}
\caption{
Quantum correlation matrix.
}
\label{apph1fig1}
\end{figure}

Figure~\ref{apph1fig1} shows the matrix element of $S$ for the ground-state quantum correlation matrix. The intensity is strong near the diagonal area, but we can still find finite amount of intensity away from the diagonal line. This is related to algebraic decay of spin correlation at quantum critical point. Because of the periodic boundary condition, we find somehow strong internsity near the upper right and lower left regions. 

\begin{figure}[htbp]
\begin{center}
\includegraphics[width=7cm]{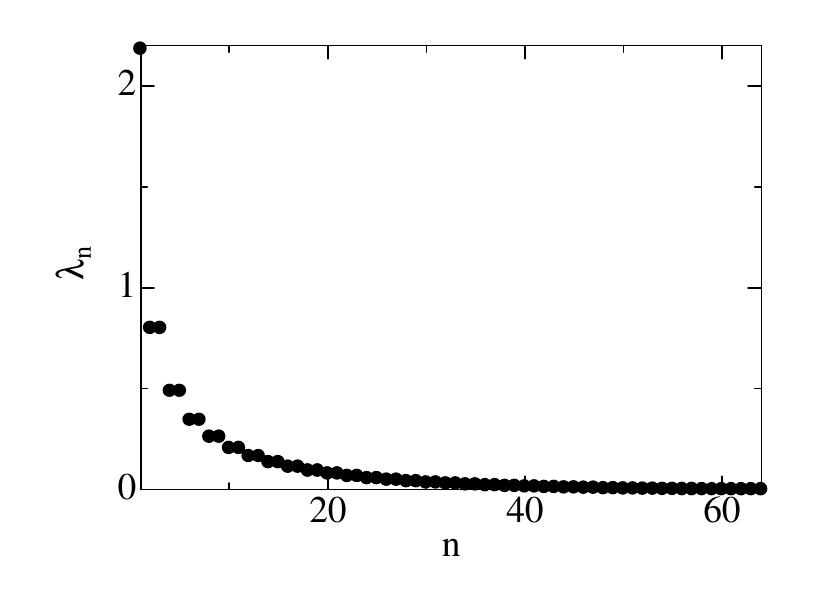}
\end{center}
\caption{
Squared singular value spectrum $\lambda_{n}$. Note that the spectrum is not normalized.
}
\label{apph1fig2}
\end{figure}

Figure~\ref{apph1fig2} shows the squared singular value spectrum. We find that the spectral data are doubly degenerate except for the largest and the smallest singular values, $\lambda_{1}$ and $\lambda_{64}$. The degeneracy is a consequence of translation of identical spin patterns as we have already discussed in the previous subsection. Only the smallest singular value $\lambda_{N=64}$ is negligible numerically, and that is consistent with Eq.~(\ref{lambdan}) in which we have $\lambda_{N=4}=0$. The envelope of the spectrum seems to show power-law decay, and this would be originated from power-law correlation of each entry in $S$ due to quantum criticality. We will discuss this point later.

\begin{figure}[htbp]
\begin{center}
\includegraphics[width=4cm]{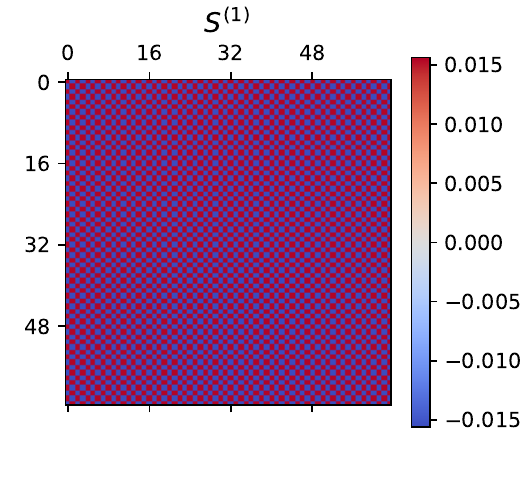}
\includegraphics[width=8cm]{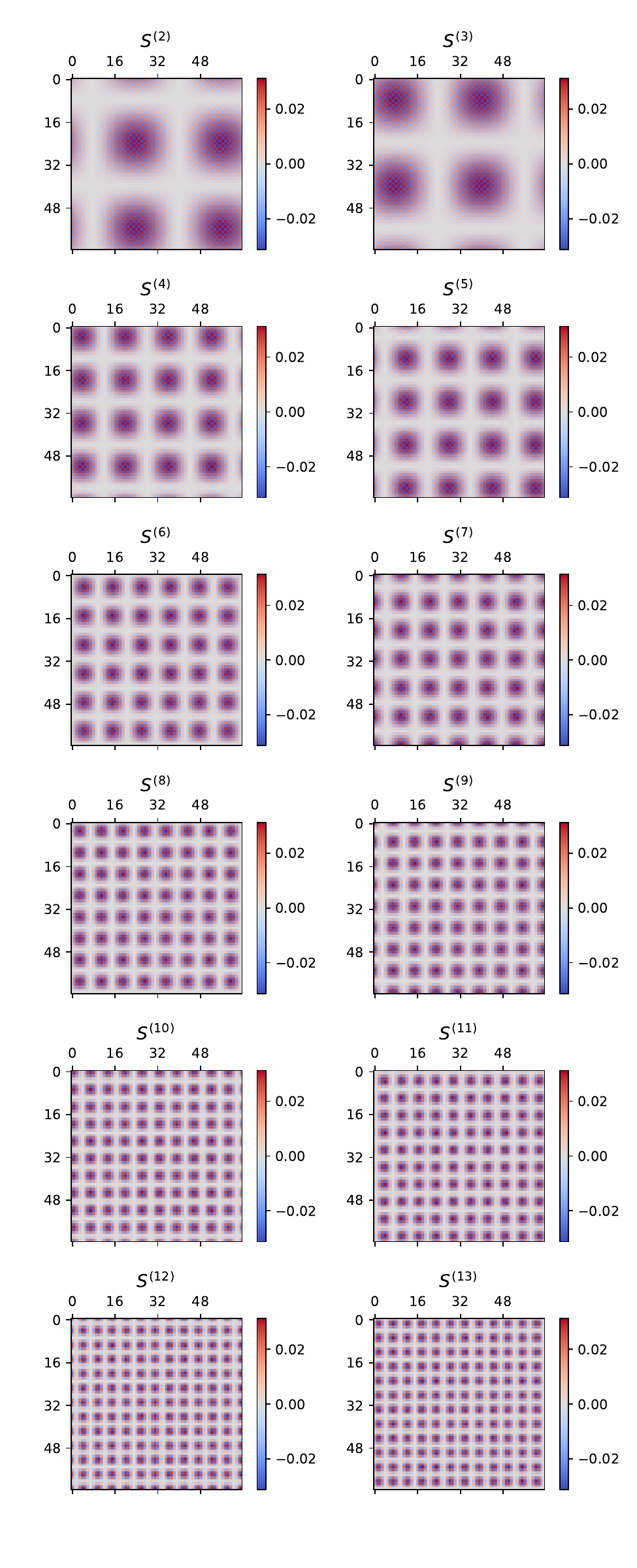}
\end{center}
\caption{
SVD components from $S^{(1)}$ to $S^{(13)}$. The data have been normalized for better presentation.
}
\label{apph1fig3}
\end{figure}

\begin{figure}[htbp]
\begin{center}
\includegraphics[width=8cm]{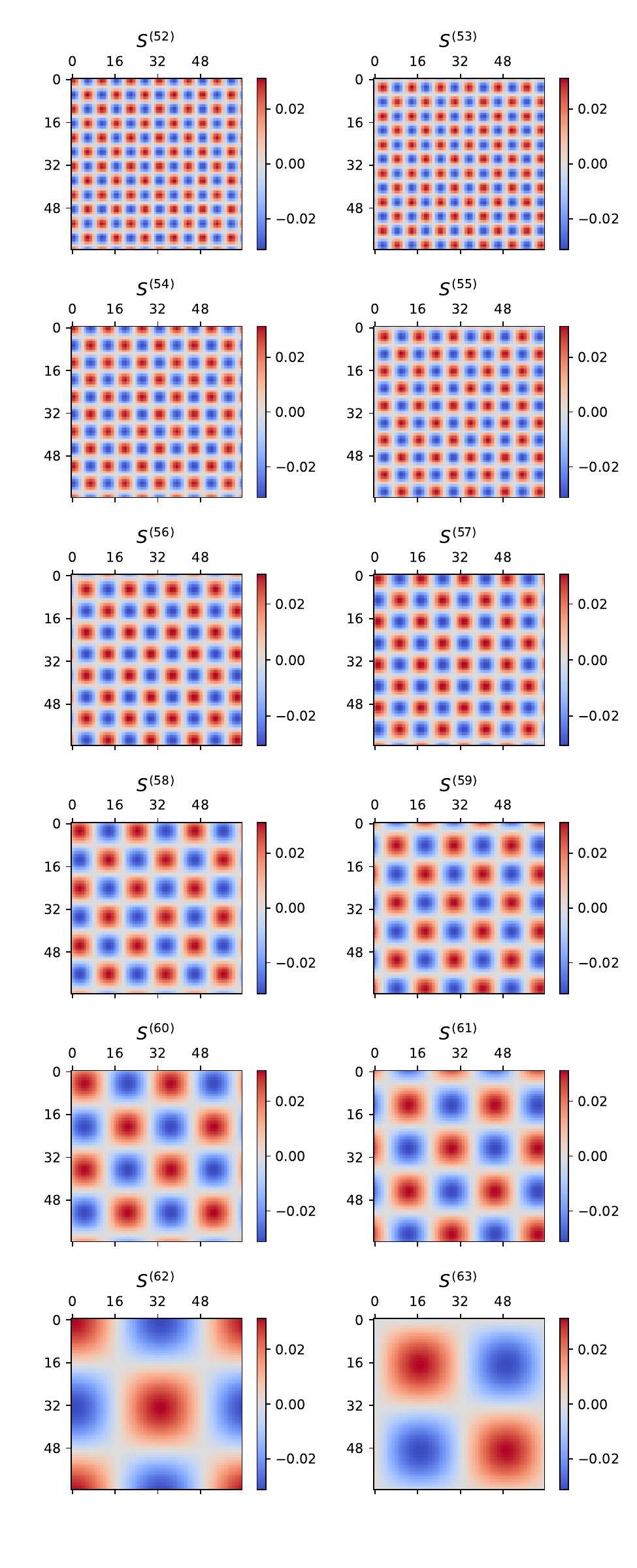}
\includegraphics[width=4cm]{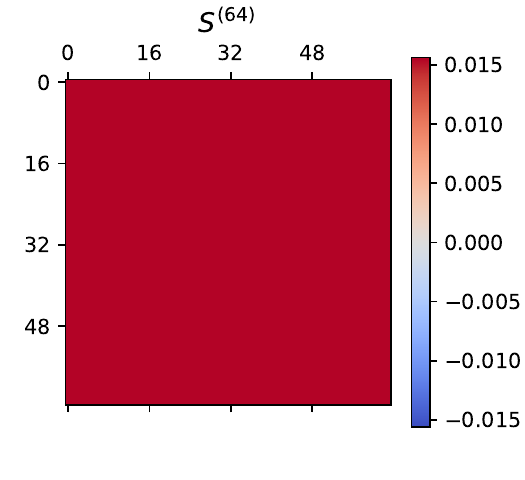}
\end{center}
\caption{
SVD components from $S^{(52)}$ to $S^{(64)}$. The data have been normalized for better presentation.
}
\label{apph1fig4}
\end{figure}

To understand the origin of the degeneracy, we show some of characteristic SVD components, $S^{(n)}$, after normalization in Fig.~\ref{apph1fig3}. These data are composed of modulated antiferromagnetic patterns (blue and red dots are mixed in each cluster). We find that $S^{(1)}$ for the largest singular value actually represent the perfect classical N\'{e}el order and each degenerate components with $n\ge 2$ represent domain excitations from the classical N\'{e}el order (Note that the domain boundaries are smeared out). The size of ordered area inside of single domain, $L_{n}$, is clearly characterized by
\begin{eqnarray}
L_{n}=\frac{N}{n}, \label{relation}
\end{eqnarray}
for $n=1$ and even $n$ values. Now we take $N=64$, and then need to assume $n<32=N/2$ in order to keep $L_{n}\ge 2$ which is the minimum size of antiferromagnetic domain. When this condition is not satisfied, we can not create antiferromagnetic domains. Because of this fact, smaller singular value components ($n\ge 32$) behave quite differently. Figure~\ref{apph1fig4} shows such tendency. In this case also, we find periodic nature of the correlation, but each cluster shows ferromagnetic correlation, not antiferromagnetic. These data represent high energy components. Except for $S^{(64)}$, we notice that the total magnetic moment is zero, since the ferromagnetic domains with different signs exist. As for $S^{(64)}$, the bulk ferromagnetic basis state still exists, but this contribution vanishes due to the fact that $\lambda_{64}=0$. Thus the condition $S_{z}^{tot}=0$ is kept successfully.

\begin{figure}[htbp]
\begin{center}
\includegraphics[width=7cm]{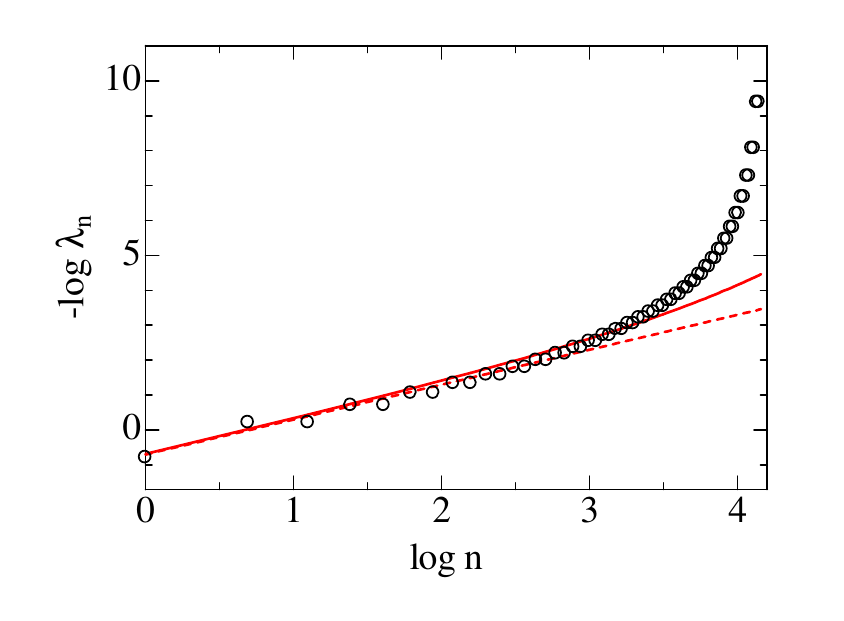}
\end{center}
\caption{
Log-log plot for the squared singular value spectrum (open circles). The red solid and dashed lines are guides for the scaling relations $\lambda_{n}\propto n^{-1}e^{-n/N}$ and $\lambda_{n}\propto n^{-1}$, respectively.
}
\label{apph1fig5}
\end{figure}

Figure~\ref{apph1fig5} shows the scaling plot for $\lambda_{n}$. This scaling corresponds to the envelope of $\lambda_{n}$ as a function of $n$ in order to fit $\lambda_{1}, \lambda_{2}, \lambda_{4}, \cdots$. For $n<N/2$ (antiferromagnetic correlation remains), we find
\begin{eqnarray}
\lambda_{n}\propto\frac{1}{n}e^{-n/N}. \label{scaling}
\end{eqnarray}
Combining Eq.~(\ref{relation}) with Eq.~(\ref{scaling}), we obtain
\begin{eqnarray}
\lambda_{n}\propto L_{n}e^{-1/L_{n}}. \label{result}
\end{eqnarray}
This is direct relationship between the SVD spectrum and domain size. When the domain size is very large, Eq.~(\ref{result}) leads to $\lambda_{n}\propto L_{n}$.

\section{Discussion}

Let us interpret the numerical results in terms of circulant matrix approach. The classical antiferromagnetic correlation is represented as
\begin{eqnarray}
\vec{X}_{N/2}\vec{X}_{N/2}^{\dagger}=\frac{1}{N}\left(\begin{array}{cccc}1&-1&1&\cdots\\-1&1&-1&\cdots\\1&-1&1&\cdots\\ \vdots&\vdots&\vdots&\ddots\end{array}\right),
\end{eqnarray}
and various domain excitations are given by
\begin{eqnarray}
\vec{X}_{k}\vec{X}_{k}^{\dagger}+\vec{X}_{N-k}\vec{X}_{N-k}^{\dagger}=\frac{2}{N}\left(\begin{array}{cccc}1&c_{1k}&c_{2k}&\cdots\\ c_{1k}&1&c_{1k}&\cdots\\ c_{2k}&c_{1k}&1&\cdots\\ \vdots&\vdots&\vdots&\ddots\end{array}\right),
\end{eqnarray}
for $k=1,2,...,N/2-1$. For zero eigenvalue, we obtain the ferromagnetic correlation
\begin{eqnarray}
\vec{X}_{N}\vec{X}_{N}^{\dagger}=\frac{1}{N}\left(\begin{array}{cccc}1&1&1&\cdots\\1&1&1&\cdots\\1&1&1&\cdots\\ \vdots&\vdots&\vdots&\ddots\end{array}\right).
\end{eqnarray}
All these basis states are consistent with the MPS results. In the standard notation of SVD, the basis set is represented as
\begin{eqnarray}
U_{i1}V_{j1}=\frac{(-1)^{|i-j|}}{N}
\end{eqnarray}
for the largest singular value and
\begin{eqnarray}
&&U_{in}V_{jn}+U_{i(n+1)}V_{j(n+1)} \nonumber \\
&&\;\;\;\; =\left(\vec{X}_{k}\vec{X}_{k}^{\dagger}+\vec{X}_{N-k}\vec{X}_{N-k}^{\dagger}\right)_{ij} \nonumber \\
&&\;\;\;\; =\frac{2}{N}\cos\left(\frac{2\pi}{N}k\left|i-j\right|\right) \nonumber \\
&&\;\;\;\; =2\frac{(-1)^{|i-j|}}{N}\cos\left(\frac{\pi}{N}n|i-j|\right),
\end{eqnarray}
for even $n$ values. For $n<N/2$, we actually find that the periodicity of the cosine function is characterized by the antiferromagnetic domain size $L_{n}=N/n$. For $n>N/2$, on the other hand, we redefine the index $n$ as $n=N-n^{\prime}$ ($n^{\prime}=1,2,...$) and find
\begin{eqnarray}
U_{in}V_{jn}+U_{i(n+1)}V_{j(n+1)}=\frac{2}{N}\cos\left(\frac{\pi}{N}n^{\prime}|i-j|\right).
\end{eqnarray}
This result also represents ferromagnetic correlation with the domain size $L_{n}^{\prime}=N/n^{\prime}$.

Let us confirm whether the algebraic decay of the correlation function $S_{ij}$ is reproduced from the scaling function estimated from MPS calculation. For this purpose, we take $S_{ij}$
\begin{eqnarray}
S_{ij}&=&\sum_{n=1}^{N}U_{in}\sqrt{\lambda_{n}}V_{jn} \nonumber \\
&=&\frac{(-1)^{|i-j|}}{N}\left[\sqrt{\lambda_{1}}+2\sum_{n}\sqrt{\lambda_{n}}\cos\left(xn\right) \right].
\end{eqnarray}
where $x=\pi|i-j|/N$ and the sum in the last line runs over $n=2,4,6,...,N-2$. According to the scaling analysis, we found $\sqrt{\lambda_{n}}\simeq\sqrt{\lambda_{1}e^{(1-n)/N}/n}$ including a normalization factor (Note that this is true only for $n<N/2$). We replace the sum with integral, and take $N\rightarrow\infty$. With the help of Laplace transformation $\int_{0}^{\infty}(1/\sqrt{t})e^{-pt}dt=\sqrt{\pi/p}$, we finally obtain
\begin{eqnarray}
S_{ij}\propto\frac{(-1)^{|i-j|}}{N}\frac{\sqrt{\sqrt{a^{2}+x^{2}}+a}}{\sqrt{a^{2}+x^{2}}}\simeq\frac{(-1)^{|i-j|}}{\sqrt{N|i-j|}},
\end{eqnarray}
where $a=1/2N$ and we assume the asymptotic regime $|i-j|\sim N$ (we should be careful for the fact that this is an ideal situation without boundary effects). The result is actually consistent with the asymptotic formula $S_{ij}\propto(-1)^{|i-j|}/|i-j|$. Thus we conclude that our scaling analysis is reasonable.

The main focus throughout this paper is to understand whether the current SVD approach is efficient for representing highly entangled quantum states like the critical ground state of the Heisenberg model. We would like to clarify this point in the final few paragraphs. As we have already mentioned in the preliminary exact analysis, the present SVD can extract information of many-body wavefunction from the correlation data. This is a nontrivial result, since the information of the wavefunction has been lost after the calculation of the correlation function. On the other hand, the Wick's theorem for fermions and bosons guarantees that any multiple-point correlators are calculated from a set of two-point correlators, and this fact supports the varidity of the correlation matrix approach presented here. Although the theorem does not hold for SU(2) spin algebra, the XY-model limit of our Hamiltonian can be transformed into spinless free fermions. Thus we believe that the correlation matrix approach is still suggestive for precisely estimating the ground-state wavefunction. Therefore, it is an intersting question whether the reconstruction of the wavefunction is realized in arbitrary $N$ cases. Now, we consider $N$-site system with $S_{z}^{tot}=0$. The matrix dimension of the Heisenberg Hamiltonian is ${}_{N}C_{N/2}\gg N$. On the other hand, the dimension of the correlation matrix is just $N$, which is much smaller than that of the Hamiltonian itself. The readers may consider that the correlation matrix approach cannot recontruct full information of the wavefunction and the preliminary result with $N=4$ is accidental. However, it does not accidentally happen. The small-size result indicates the presence of some mechanism for efficient data compression.

\begin{table}[htbp]
\begin{center}
\caption{Ground-state data for $N=8$ Heisenberg model}
\begin{tabular}{c|c|c|c|c}
$l$&$c_{l}$&$d_{l}$&$\left|r_{l}\right>$&$q_{l}/\pi$ \\ \hline
$1$ & $0.398297$ & $2$ & $\left|\uparrow\downarrow\uparrow\downarrow\uparrow\downarrow\uparrow\downarrow\right>$ & $1.0$  \\ \hline
$2$ & $-0.164406$ & $16$ & $\left|\downarrow\uparrow\uparrow\downarrow\uparrow\downarrow\uparrow\downarrow\right>$ & $0.86$  \\ \hline
$3$ & $0.137134$ & $8$ & $\left|\uparrow\downarrow\uparrow\downarrow\downarrow\uparrow\downarrow\uparrow\right>$ & $0.81$  \\ \hline
$4$ & $0.090059$ & $4$ & $\left|\uparrow\downarrow\downarrow\uparrow\uparrow\downarrow\downarrow\uparrow\right>$ & $0.53$  \\ \hline
$5$ & $0.054545$ & $16$ & $\left|\uparrow\uparrow\downarrow\downarrow\uparrow\downarrow\downarrow\uparrow\right>$ & $0.59$  \\ \hline
$6$ & $-0.034742$ & $16$ & $\left|\uparrow\uparrow\uparrow\downarrow\uparrow\downarrow\downarrow\downarrow\right>$ & $0.17$ \\ \hline
$7$ & $0.00747$ & $8$ & $\left|\uparrow\uparrow\uparrow\uparrow\downarrow\downarrow\downarrow\downarrow\right>$ & $0.19$
\end{tabular}
\label{tab}
\end{center}
\end{table}

To see the data compression mechanism, let us look at the case of $N=8$. The matrix dimension of the Hamiltonian is ${}_{8}C_{4}=70\gg 8$. By the exact diagonalization of the Hamiltonian matrix, we find that the ground state is represented by
\begin{eqnarray}
\left|\psi\right>=\sum_{l=1}^{7}c_{l}\left|\phi_{l}\right>=\sum_{l=1}^{7}c_{l}\sum_{i=1}^{d_{l}}G_{li}\left|r_{l}\right>.
\end{eqnarray}
Here, the ground state is classified into $7$ groups. Each group labeled by $l$ has a reference state $\left|r_{l}\right>$, and the other states in the group $l$ are generated by space translation and global spin inversion from the reference state. The operator $G_{li}$ represents such translation and inversion procedures, and $d_{l}$ is the total number of states in the group $l$. Note that $d_{l}$ is small for high symmetry cases. The ground-state data are summarized in Table~\ref{tab}, where $\sum_{l=1}^{7}c_{l}^{2}d_{l}=1$ and $\sum_{l=1}^{7}d_{l}=70$. The coefficients $\{c_{l}\}$ are in descending order, and in this case the order is closely related to the size of the magnetic domain like the previous Fourier analysis. Furthermore, the trend of $c_{l}$ seems to be related to the scaling formula in Eq.~(\ref{scaling}). The reason for these tendencies should be clarified in the following sentences. In Table~\ref{tab}, we calculate $f_{l}(q)=\sum_{i=1}^{8}\sum_{j=1}^{8}\left<r_{l}\right|S_{i}^{z}S_{j}^{z}\left|r_{l}\right>\cos\left(q\left|i-j\right|\right)$, and evaluate the $q$ value, $q_{l}$, that gives the maximum of $f_{l}(q)$. The quantity $q_{l}$ roughly estimates periodicity of the domain structure. We find that $\left|\phi_{1}\right>$ is the bulk antiferromagnetic configuration, $\left|\phi_{1}\right>=\left|\uparrow\downarrow\uparrow\downarrow\uparrow\downarrow\uparrow\downarrow\right>+\left|\downarrow\uparrow\downarrow\uparrow\downarrow\uparrow\downarrow\uparrow\right>$ with periodicity $\pi$ and
\begin{eqnarray}
c_{1}^{2}\left<\phi_{1}\right|S_{i}^{z}S_{j}^{z}\left|\phi_{1}\right>\simeq\chi_{4}\left(\vec{X}_{4}\vec{X}_{4}^{\dagger}\right)_{ij}=\sqrt{\lambda_{1}}U_{i1}V_{j1}.
\end{eqnarray}
Note that this is an approximated result for $N=8$, since there is a bit of a discrepancy between the spin (digital) and Fourier (analog) basis approaches. As we increase the number $l$, smaller-domain states appear sequentially, and the correlation tends to change into ferromagnetic one. As for $q_{l}$ and $c_{l}$, we find the following rough tendencies
\begin{eqnarray}
q_{1}>q_{2}\simeq q_{3}>q_{4}\simeq q_{5}>q_{6}\simeq q_{7},
\end{eqnarray}
and
\begin{eqnarray}
\left|c_{1}\right|>\left|c_{2}\right|\sim\left|c_{3}\right|>\left|c_{4}\right|\sim\left|c_{5}\right|>\left|c_{6}\right|\sim\left|c_{7}\right|.
\end{eqnarray}
These results show that the states are categorized into $4 (=N/2)$ different domain structures. For instance, let us focus on $\left|\phi_{2}\right>$ and $\left|\phi_{3}\right>$ that have almost the same $q_{l}$ value. We consider that they are in the same category. We find that both of these states contain two antiferromagnetic domains. The state $\left|\phi_{3}\right>$ shows regular domain patterns ($\uparrow\downarrow\uparrow\downarrow$ and $\downarrow\uparrow\downarrow\uparrow$), but $\left|\phi_{2}\right>$ irregular ($\downarrow\uparrow$ and $\uparrow\downarrow\uparrow\downarrow\uparrow\downarrow$). In other words, each cetegory is composed of a set of states in which the domain boundaries are fluctuating. Even if we consider the irregular patterns, the periodicity still holds, since we sum up all possible states generated by translation and inversion. When we define $\left|\varphi\right>=c_{2}\left|\phi_{2}\right>+c_{3}\left|\phi_{3}\right>$ and calculate $S_{ij}(\varphi)=\left<\varphi\right|S_{i}^{z}S_{j}^{z}\left|\varphi\right>$, we find that the Fourier transform of $S_{ij}(\varphi)$ has a peak at $3\pi/4$ and
\begin{eqnarray}
S_{ij}(\varphi)&\simeq&\chi_{3}\left(\vec{X}_{3}\vec{X}_{3}^{\dagger}+\vec{X}_{5}\vec{X}_{5}^{\dagger}\right)_{ij} \nonumber \\
&=&\sqrt{\lambda_{2}}\left(U_{i2}V_{j2}+U_{i3}V_{j3}\right).
\end{eqnarray}
The same discussion is possible for the state pair, $\left|\phi_{4}\right>$ and $\left|\phi_{5}\right>$ ($\left|\phi_{6}\right>$ and $\left|\phi_{7}\right>$). Therefore, the wavefunction estimation can be almost realized within the condition that the states in the same category cannot be separately treated due to the formation of smooth boundary states. In the $4$-site case, the domain is too small, and there is no boundary fluctuation. Thus the simple classification was possible. In numerical simulation for $N=64$, the domain boundaries are smeared out, and the loss of sharp domain edge is due to the mixture of regular and irregular domain bases.

The remaining task to complete answer for arbitraly $N$ is a combinatorial approach in which we count the number of microstates that belong to each category. The Hamiltonian dimension is given by ${}_{N}C_{N/2}$, and the average of the number of states in each category is estimated as ${}_{N}C_{N/2}/(N/2)$. The number of states in each group is maximally $2N$ (translation and inversion), and thus the number of different reference states (the number of groups) in the same category is roughly given by ${}_{N}C_{N/2}/N^{2}$. To make a smooth cosine function, we need to consider spin fluctuation near the domain boundaries. If we take $x$ as an average number of spins that participate in the boundary fluctuation, $x$ is estimated as $2^{x}\sim{}_{N}C_{N/2}/N^{2}$. For instance, we obtain $x\sim N-2\log_{2}N=52<N$ for $N=64$, where we used the Stirling formula for the evaluation of the combination. The result is actually comparable to Figs.~\ref{apph1fig3} and~\ref{apph1fig4}. In the large-$N$ limit, the deviation of discrete spin basis from the smooth Fourier basis becomes minimum.

Let us summarize our discussion. The ground state is represented by
\begin{eqnarray}
\left|\psi\right>=\sum_{m=1}^{N/2}\left|\varphi_{m}\right>=\sum_{m=1}^{N/2}\sum_{l\in C_{m}}c_{l}\left|\phi_{l}\right>,
\end{eqnarray}
where the index $m$ represents each category of domain structure and $C_{m}$ is a set of integers that determines the same category. The number of the integers has been estimated as $x$ in the previous paragraph. Our reasonable conjectures in arbitrary $N$ cases are
\begin{eqnarray}
S_{ij}(\varphi_{1})=c_{1}^{2}\left<\phi_{1}\right|S_{i}^{z}S_{j}^{z}\left|\phi_{1}\right>=\sqrt{\lambda_{1}}U_{i1}V_{j1}, \label{final1}
\end{eqnarray}
and
\begin{eqnarray}
S_{ij}(\varphi_{m})&=&\sum_{l\in C_{m}}c_{l}^{2}\left<\phi_{l}\right|S_{i}^{z}S_{j}^{z}\left|\phi_{l}\right> \nonumber \\
&=&\sqrt{\lambda_{2m-2}}\left\{U_{j(2m-2)}V_{j(2m-2)} \right. \nonumber \\
&&\;\;\;\;\;\;\;\;\;\;\;\;\;\;\;\; \left. +U_{i(2m-1)}V_{j(2m-1)}\right\}, \label{final2}
\end{eqnarray}
for $m=2,3,...,N/2$. The result gives close relationship between the singular value spectrum and the weights of the basis states, and thus the precise estimation of the wavefunction is really possible. Both the Fourier and domain basis are determined by hand so that they match with each other, and thus the real physical information is contained only in $\sqrt{\lambda_{n}}$ and $c_{l}$. This quite natural procedure originates in the derivation of Eqs.~(\ref{final1}) and (\ref{final2}).

\section{Summary and Future Perspective}

We presented SVD of correlation matrix of 1D antiferromagnetic quantum Heisenberg model. We found that the decomposition naturally creates a hierarchy of data set that coinsides with the ground and various domain excitations in the classical 1D antiferromagnetic model. The SVD analysis for the correlation matrix gives us a very natural way of a quantum PCA in the sense that the SVD gives us important information of the ground-state wavefunction, and overcomes disadvantage of treating local entanglement. The classical limit of the ground state is the principle component, but at the same time the domain excitations play important roles in the residual singular values. We have obtained important relationships associated with domain size, the scaling of SVD spectrum, and reconstruction formula of the two-point correlator.

The presence of a successful method for selectively observing basis states of a quantum wavefunction may also be suggestive for the study of weak value, weak measurement, and related topics associated with quantum measurement that does not break quantum superposition~\cite{Aharonov1,Aharonov2,Brodutch}.

The approach presented in this paper can be applicable to any quantum many-body models, some of which may not have clear order parameters, finite-temperature cases, and time evolution to examine the thermal pure state approach for development of fundamentals of statistical mechanics. Thus, this flexibility would lead to generic approach for construction of quantum PCA. We also hope that the present analysis opens a new door to have deeper understanding for entanglement. There are still many open issues to be resolved.

This work was supported by JSPS KAKENHI Grant Number 18K03474. HM acknowledges Yoichiro Hashizume and Kunio Ishida for fruitful discussion. HM also acknowledges Yusuke Masaki for critical reading of the manuscript.


\begin{thebibliography}{99}

\bibitem{White1}
Steven R. White, Phys. Rev. Lett. {\bf 69}, 2863 (1992).
\bibitem{White2}
Steven R. White, Phys. Rev. B {\bf 48}, 10345 (1993).
\bibitem{Cirac}
F. Verstraete, D. Porras, and J. I. Cirac, Phys. Rev. Lett. {\bf 93}, 227205 (2004).
\bibitem{Lloyd1}
Seth Lloyd, Masoud Mohseni, and Patrick Rebentrost, Nat. Phys. {\bf 10}, 631 (2014).
\bibitem{Mosetti}
Renzo Mosetti, arXiv:1503.00872.
\bibitem{Lloyd2}
Patrick Rebentrost, Adrian Steffens, Iman Marvian, and Seth Lloyd, Phys. Rev. A {\bf 97}, 012327 (2018).
\bibitem{Tang}
Ewin Tang, arXiv:1811.00414.
\bibitem{Ostlund1}
S. $\ddot{\rm O}$stlund and S. Rommer, Phys. Rev. Lett. {\bf 75}, 3537 (1995).
\bibitem{Ostlund2}
S. Rommer and S. $\ddot{\rm O}$stlund, Phys. Rev. B {\bf 55}, 2164 (1997).
\bibitem{Verstraete1}
F. Verstraete, D. Parras, and J. I. Cirac, Phys. Rev. Lett. {\bf 93}, 227205 (2004).
\bibitem{Verstraete2}
F. Verstraete and J. I. Cirac, arXiv:0407066..
\bibitem{Vidal}
Guifre Vidal, Phys. Rev. Lett. {\bf 99}, 220405 (2007).
\bibitem{Evenbly1}
G. Evenbly and G. Vidal, Phys. Rev. B {\bf 79}, 144108 (2009).
\bibitem{Matsueda1}
H. Matsueda, Phys. Rev. E {\bf 85}, 031101 (2012).
\bibitem{Matsueda2}
Ching Hua Lee, Yuki Yamada, Tatsuya Kumamoto, and Hiroaki Matsueda, J. Phys. Soc. Jpn. {\bf 84}, 013001 (2015).
\bibitem{Okunishi1}
Y. Imura, T. Okubo, S. Morita, and K. Okunishi, J. Phys. Soc. Jpn. {\bf 83}, 114002 (2014).
\bibitem{Matsueda3}
Hiroaki Matsueda, Ching Hua Lee, and Yoichiro Hashizume, J. Phys. Soc. Jpn. {\bf 85}, 086001 (2016).
\bibitem{Ozaki1}
Hiroaki Matsueda and Dai Ozaki, Phys. Rev. E {\bf 92}, 042167 (2015).
\bibitem{Ozaki2}
Ching Hua Lee, Dai Ozaki, and Hiroaki Matsueda, Phys. Rev. E {\bf 94}, 062144 (2016).
\bibitem{Henley1}
Siew-Ann Cheong and Christopher L. Henley, Phys. Rev. B {\bf 69}, 075111 (2004).
\bibitem{Henley2}
Siew-Ann Cheong and Christopher L. Henley, Phys. Rev. B {\bf 69}, 075112 (2004).
\bibitem{XiaoLiang}
Xiao-Liang Qi, arXiv:1309.6282.
\bibitem{ChingHua}
Ching Hua Lee and Xiao-Liang Qi, Phys. Rev. B {\bf 93}, 035112 (2016).
\bibitem{Aharonov1}
Yakir Aharonov, David Z. Albert, and Lev Vaidman, Phys. Rev. Lett. {\bf 60}, 1351 (1988).
\bibitem{Aharonov2}
Yakir Aharonov and Lev Vaidman, Phys. Rev. A {\bf 41}, 11 (1990).
\bibitem{Brodutch}
Aharon Brodutch and Eliahu Cohen, Phys. Rev. Lett. {\bf 116}, 070404 (2016).

\end{thebibliography}
\end{document}